\documentclass[12pt]{article}

\usepackage{graphicx}
\usepackage{epsfig}
\usepackage{psfrag}
\usepackage{wrapfig}
\usepackage[all]{xy}

\usepackage{fullpage}
\usepackage{setspace}
\usepackage{mathptmx}
\usepackage{url}
\usepackage{booktabs}

\usepackage{footmisc}
\newcommand\wordcount{
    \immediate\write18{texcount -sum -1 \jobname.tex > 'count.txt'}
\input{count.txt}words}

\usepackage{natbib}

\usepackage{amsmath}
\usepackage{amsfonts}
\usepackage{amsthm}
\usepackage{bbm}
\usepackage{array}

\def\Var{{\rm Var}\,}
\def\E{{\rm E}\,}

\newcommand{\arrowp}{\stackrel{p}{\rightarrow}}

\newcommand{\RR}{\mathbb{R}}

\begin{document}

\sloppy

\title {A Note on ``How Robust Standard Errors Expose Methodological Problems They Do Not Fix, and What to Do About It"}
%
%

\author{Peter M. Aronow\thanks{Peter M. Aronow is Assistant Professor, Departments of Political Science and Biostatistics, Yale University, 77 Prospect St., New Haven, CT 06520 (Email: peter.aronow@yale.edu). The author thanks Alex Coppock, Gary King, Winston Lin, Helen Milner, Eric Neumayer, Molly Offer-Westort, David Pollard, and Molly Roberts for helpful discussions.} }


\maketitle

\begin{abstract}
\citet[KR]{kingroberts2015} claim that a disagreement between robust and classical standard errors exposes model misspecification. We emphasize that KR's claim only generally applies to parametric models: models that assume a restrictive form of the distribution of the outcome. Many common models in use in political science, including the linear model, are not necessarily parametric -- rather they may be semiparametric. Common estimators of model parameters such as ordinary least squares have both robust (corresponding to a semiparametric model) and classical (corresponding to a more restrictive model) standard error estimates. Given a properly specified semiparametric model and mild regularity conditions, the classical standard errors are not generally consistent, but the robust standard errors are. To illustrate this point, we consider the case of the regression estimate of a semiparametric linear model with no model misspecification, and show that robust standard errors may nevertheless systematically differ from classical standard errors. We show that a disagreement between robust and classical standard errors is not generally suitable as a diagnostic for regression estimators, and that KR's reanalyses of \citet{neumayer2003} and \citet{buthemilner2008} are predicated on strong assumptions that the original authors did not invoke nor require.
\end{abstract}

\section{Introduction}

\citeauthor{kingroberts2015} (henceforth KR), published a provocative article, ``How Robust Standard Errors Expose Methodological Problems They Do Not Fix, and What to Do About It," in {\it Political Analysis} in 2015. Echoing insights from \citet{freedman2006}, KR claim that a disagreement between robust and classical standard errors indicates model misspecification. KR proceed to argue that this disagreement should be corrected via respecification of the model. In their introduction, KR state that ``robust and classical standard errors that differ need to be seen as bright red flags that signal compelling evidence of uncorrected model misspecification. They highlight statistical analyses begging to be replicated, respecified, and reanalyzed, and conclusions that may need serious revision'' (p. 159).

However, KR restrict their attention to parametric models, or models that impose very strong distributional assumptions. Loosely speaking, parametric models require the assumption that the distribution of the outcome is known to the researcher up to a finite-dimensional parameter vector. Models that do not make this assumption are known as {\it semiparametric}. (Neither the term ``semiparametric" nor the closely related term ``nonparametric" are used at any point in KR.) In footnote 3 (and also, to a lesser extent on pp. 160, 162, 164 and 165), KR acknowledge the limitations of their setting, referring to alternative approaches as ``forgo[ing] models" or ``non-model-based." While KR suggest these qualifications to their claims throughout the manuscript, in our view, these important scope conditions are underemphasized and easily missed by a casual reader. KR casts doubt on the utility of models that are not fully parametric, stating that the researcher ``explicitly gives up the ability to compute most quantities from the model in return for the possibility of valid inference for some." This logic further informs KR's suggestion of a parametric bootstrap-based information matrix test \citep{dhaenehoorelbeke2004} as a post-estimation diagnostic.

We agree with KR that a disagreement between robust standard errors and classical standard errors in the context of parametric models indicates model misspecification that should be investigated. However, the restriction to parametric models implies a set of scope conditions that excludes a large body of empirical research and commonly used models. Semiparametric models are broadly applied in situations where the researcher does not have theoretical justification for a set of distributional assumptions about, e.g., the shape of the outcome. For example, as we will discuss in Section 1.2, textbook treatments of the linear model and ordinary least squares estimation are frequently motivated by semiparametric modeling assumptions. As \citet{davidsonmackinnon2004} note, ``Both [semiparametric and parametric] models are frequently encountered in econometrics." The restrictive scope conditions of KR's argument exclude the very work that KR critique. KR raise alarm at the fact that 66\% of papers using regression analyses published in the {\it American Political Science Review} between 2009-2012 report robust standard errors at some point, but do not consider that many of the models employed have a semiparametric interpretation. In fact, two of the three articles that KR reanalyze and purport to correct fall outside of the scope conditions implied by a restriction to parametric models. 
 
Our note proceeds as follows. In Sections 1.1 and 1.2, we provide important definitions and discuss common linear models that are (infinitely) less restrictive than KR's linear-normal model.  In Section 1.3, we consider KR's reanalyses of \citet{neumayer2003} and \citet{buthemilner2008}, and show that KR's diagnostics are not applicable. In Section 1.4, we discuss the role of theory in model specification, and argue that KR conflate an unspecified feature of a model with a misspecified feature of a model. Section 2 provides an illustrative example of a properly specified semiparametric linear model for which the robust and classical standard errors disagree. In Section 3, we discuss the broader implications of KR's recommendations for quantitative inquiry.
 
\subsection{Parametric and Semiparametric Models}

An extremely intuitive definition of parametric and semiparametric models is given by \citet{kennedyinpress}. (We nearly quote Kennedy verbatim in the next two sentences, adapting the language slightly for context.) A statistical model is a set of possible probability distributions, which is assumed to contain the distribution of observable data. In a parametric model, the model is assumed to be indexed by a {\it finite-dimensional} real-valued parameter vector that lies in $\RR^q$, where $q$ is finite. Semiparametric models are sets of probability distributions that {\it cannot} be indexed by only a finite-dimensional real-valued parameter, i.e., models that are indexed by an infinite-dimensional parameter. Both parametric and semiparametric models are models, however they vary in how much they assume about the universe. Put simply: any parametric model assumes infinitely more about the process that gives rise to the data than does any semiparametric model that nests it. Whenever, e.g., the full distribution of a disturbance term in a model is left unspecified, it implies that the researcher admits a wide (in fact infinite) range of possibilities about its exact distribution.

A useful concept in the study of semiparametric models is the {\it parametric submodel}. Again paraphrasing Kennedy, a parametric submodel of a semiparametric model is a parametric model that is (i) contained in the semiparametric model and (ii) contains the truth. When a parametric model is correct, it can be represented as a parametric submodel of some greater semiparametric model. For example, among the (infinite) set of models for a single variable $X_i$ such that $\E[X_i] = 0$, the parametric model implied by $X_i \sim N(0,\sigma^2)$ is one potential submodel. If indeed $X_i$ is distributed normally, then this parametric model is indeed a parametric submodel. As we proceed, we will see how classical standard errors are correct under the maintained assumption of a given parametric submodel, but they are not generally correct when that maintained assumption is incorrect.

\subsection{Linear Models and Ordinary Least Squares Regression}

One particularly common semiparametric model is the linear model. (Details on semiparametric specification of the linear model are provided in Section 2.1.) Given its prevalence in applied research \citep{angristpischke2009}, the linear model (and ordinary least squares estimation thereof), is the focus of our note. Ordinary least squares estimation can be motivated by both semiparametric and parametric linear models. The robust standard errors generally correspond to a semiparametric model that puts only limited restrictions on the disturbance distribution. The use of ordinary least squares does not imply a commitment to any specific parametric model, nor even to semiparametric models with homoskedastic disturbances. 

In their discussion of the linear model, KR focus on one special case: the classical linear-normal model, which assumes homoskedastic normal disturbances. However, textbook discussions of the linear model are often motivated without an assumption of normal disturbances (\citealt[Ch. 4]{wooldridge2002}; \citealt[Ch. 14]{rice2007}; \citealt[Ch. 13]{wasserman2004}; \citealt[Ch. 4]{freedman}; \citealt[Ch. 15-16]{goldberger1991}) and further without homoskedasticity (\citealt[Ch. 2]{wooldridge2000}; 
\citealt[Ch. 4]{hansen2000}; \citealt[Ch. 27]{goldberger1991}; \citealt[Ch. 4]{camerontrivedi2005}; 
\citealt[Ch. 6]{stockwatson2011}; \citealt[Ch. 2.3]{hayashi2000}). Although further specification of the disturbance distribution can improve efficiency (justifying, e.g., the semiparametric motivation for weighted least squares, or maximum likelihood estimation with parametric specification), it is not necessary for model identification, and misspecification may lead to inconsistency. In fact, the classical ``ideal" linear model where the Gauss-Markov assumptions hold is semiparametric, as the full distribution of the disturbance term is left unspecified.\footnote{Recall that the Gauss-Markov theorem does not require normality of the disturbances nor of any other features of the model. Loosely speaking, the Gauss-Markov assumptions are linearity, noncollinearity, and uncorrelated, mean-zero, homoskedastic errors.} Although classical and robust standard errors asymptotically agree under the Gauss-Markov assumptions, KR's argument implies that the Gauss-Markov assumptions alone are not strong enough to specify a model. The parametric bootstrap-based information matrix test advocated by KR cannot generally be used for linear models that satisfy the Gauss-Markov assumptions, but for which the researcher has not specified the full distribution of the disturbance term.

\subsection{Applications to \citet{neumayer2003} and \citet{buthemilner2008}}

KR's caveat that their approach is only applicable to parametric models is not addressed in their own applications. KR apply their recommendations to two empirical studies that invoke a linear model: \citet{neumayer2003} and \citet{buthemilner2008}. (KR also apply their recommendations to \citet{dreherjensen2007} who use a parametric Poisson model, and thus the diagnostics -- but not necessarily the recommendations -- are not inappropriate.) We consider each of these applications in turn.

\citet{neumayer2003} considers the predictors of aid allocation by multilateral donors. The estimation strategy is described as: ``All regressions are run with pooled ordinary least squares (OLS) and heteroscedasticity and serial-correlation robust standard errors" (p. 109). Neumayer does not assume normal errors, and explicitly does not assume homoskedasticity. KR, in their reanalysis of the results, presume that having homoskedastic normal errors constitutes the appropriate benchmark model. But these are not the assumptions that the original author invoked, nor are they necessarily justifiable by theory. Beyond their examination of robust vs. classical standard errors, KR apply an information matrix test against the null hypothesis of homoskedastic normal errors, and further present Q-Q plots of the residuals against a normal distribution. These are all inappropriate as diagnostics of Neumayer's approach -- KR are evaluating an estimation approach justifiable by semiparametric assumptions by the standard of an unnecessary parametric model.

\citet{buthemilner2008} consider the relationship between foreign direct investment and international trade agreements. Like \citet{neumayer2003}, \citet{buthemilner2008} invoke no assumptions about homoskedasticity or normal errors. In fact, \citet{buthemilner2008} rather explicitly are estimating a semiparametric model with potentially heteroskedastic errors (p. 747). The fact that \citet{buthemilner2008} are using fixed effects estimation with a short panel implies that their model is imbued with a semiparametric interpretation, as the structure of the conditional expectation function is left partially unspecified. Again, KR's reanalysis of \citet{buthemilner2008} is predicated on assumptions that the original authors did not make. KR justify their investigation in part by stating that the difference between the robust and classical standard errors ``could suggest a significant amount of heteroskedasticity in the data, which at best indicates inefficiency for some quantities and bias in others, or it could suggest model misspecification that biases all relevant quantities" (p. 173). The efficiency concern is addressed directly by \citet{buthemilner2008}'s generalized least squares specification, but this specification -- included in the main text (Table 4) -- is not noted by KR. The ``bias" in other quantities may not be relevant at all, as it may apply to quantities that \citet{buthemilner2008} do not estimate. Finally, the idea that KR's diagnostics indicate model misspecification is only true if we were to believe that heteroskedasticity indicates misspecification of relevant elements of the conditional expectation function, which has no general theoretical basis.\footnote{There exist tests of unmodelled nonlinearity in the conditional expectation function -- e.g., \citet{ramsey,hainmueller} -- and if such nonlinearity is of scientific interest then such tests are more appropriate than are KR's proposals. To paraphrase John Roberts \citep{roberts}, our view is that the way to test for nonlinearity in the conditional expectation function is to test for nonlinearity in the conditional expectation function.} Thus, as with \citet{neumayer2003}, KR's reanalysis of \citet{buthemilner2008} is only justified if KR's assumptions are met -- assumptions that the original authors did not make.

\subsection{Theory and Specification}

Throughout the manuscript, KR characterize robust standard errors as being used ``to correct standard errors for model misspecification" (p. 159). On a fundamental level, KR conflate an unspecified feature of a model with a misspecified feature of a model. KR state that ``if robust and classical standard errors diverge---which means the author acknowledges that one part of his or her model is wrong---then why should readers believe that all the other parts of the model that have not been examined are correctly specified" (p. 160). If classical and robust standard errors differ for semiparametrically identified models, it may only indicate that an assumption not invoked by the researcher does not hold. Robust standard errors may be motivated by a researcher's invocation of a semiparametric model, for which the exact form of the variance of the disturbance term is not specified. As \citet[p. 131]{hayashi2000} states, ``With the advent of robust standard errors allowing us to do inference without specifying the conditional second moment ... testing conditional homoskedasticity is not as important as it used to be."\footnote{\citet[Section 9.3]{hansen2000} goes farther: ``tests for heteroskedasticity should be used to answer the scientific question of whether or not the conditional variance is a function of the regressors. If this question is not of economic interest, then there is no value in conducting a test for heteroskedasticity."} Note that this is not ``when misspecifying the conditional second moment" -- it is {\it without specifying}.

A semiparametric model is not just a parametric model that allows for misspecification; rather it is a broader model that admits the researcher's ignorance. That is to say, a theoretical model may not give enough structure to assume the entire distribution of the disturbance. The researcher's reluctance to invoke arbitrary modeling assumptions for these components should not be viewed as an admission that the model is wrong, but rather should be seen as reflecting the possibility that the researcher believes that the model includes components that may not lend themselves to parametric models. There are many cases where researchers would feel comfortable given a finite-dimensional parametrization of, e.g., elements of the conditional expectation function, but leave the remainder of the distribution unspecified. The most obvious case is when the predictors of interest have a finite, countable set of unique values, and the remainder of the predictors are incorporated into nuisance functions -- for an empirical example, see, e.g., \citet{cohenetal2015}. Similarly, researchers may have theoretical justification for assuming a particular functional form of the conditional expectation function but not the conditional variance function, see, e.g., \citet{allcotttaubinsky2015} for such a theoretical model. In such cases, there is no general justification for further specification of the disturbance term.


\section{An Example}

As we proceed, we will consider the special case of a simple linear model with a binary predictor, largely to ease exposition. We will show that: the model parameters are identified; ordinary least squares regression is consistent for the model parameters; and the robust standard errors differ from the classical standard errors (even asymptotically) without any model misspecification.

\subsection{Model and Identification}

We consider the following simple linear model: 
\begin{equation}\label{model}
Y_i = \beta_0 + \beta_1 X_i + e_i,
\end{equation} where $Y_i$ is an outcome variable; $X_i$ is a binary predictor; $e_i$ is a random disturbance term; and $\beta_0$ and $\beta_1$ are model coefficients. We follow econometric convention and condition on the $X_i$ values. Our results do not depend on this, and hold with i.i.d. sampling from $(Y_i,X_i)$, although the precise statement of our assumptions would need to be adjusted accordingly. Assume an index ordering such that $X_1,...,X_m = 1$ and $X_{m+1},...,X_n = 0$, where $m = \sum_{i=1}^n X_i$ with $0 < m < n$. 

To identify the model parameters, we impose the following semiparametric assumptions: First, assume mutual independence of the disturbances: $\forall A \subseteq \{1,2,...,n\}$, $\forall (e_1,e_2,...,e_n) \in \RR^n$, $F_{A}((e_i)_{i\in A}) = \prod_{i \in A} F_i(e_i)$. 
This is a standard assumption, and what is typically meant by ``independent" in the context of ``independent and identically distributed" disturbances. We do not assume that the disturbances are identically distributed, however. Second, assume exogeneity of $X_i$, so that $\E[e_i] = 0$ and $\E[e_i X_i] = 0$, $\forall i$. Third, assume that the following regularity conditions hold: (i) non-collinearity: $\frac{1}{n}\sum_{i=1}^n X_i \rightarrow \mu \in (0,1)$ and (ii) uniformly bounded fourth moments on the disturbance distribution: $\sup_{i \in \{1,...,n\}} \E[e_i^4] < \infty$. The classical linear-normal model considered by KR includes all of these assumptions, but adds the additional assumption that the disturbances are normally distributed and homoskedastic: $e_i \sim N(0,\sigma^2), \forall i$. In contrast, we have {\it not} specified the full probability distribution of $e_i$, nor have we invoked an assumption of homoskedasticity. Model identification follows principally from the exogeneity conditions $\E[e_i] = 0$ and $\E[e_i X_i] = 0, \forall i$.

\subsection{Ordinary Least Squares Estimation}

We need no further assumptions to derive the ordinary least squares estimate $(\hat \beta_0, \hat \beta_1)$ as a consistent estimator for $(\beta_0, \beta_1)$ using the method of moments. The identifying assumptions are equivalent to $\E[Y_i - \beta_0 - \beta_1 X_i] = 0$ and $\E[(Y_i - \beta_0 - \beta_1 X_i) X_i] = 0$. Solving the system of equations, we have $\beta_0 = \E[Y_i | X_i = 0]$ and $\beta_1 = \E[Y_i | X_i = 1] - \E[Y_i | X_i = 0]$. This yields the natural sample analogue estimators (equivalent to the ordinary least squares estimators),
\begin{equation*}\label{ols}
\hat \beta_0 = \frac{\sum_{i=m+1}^n Y_i}{n-m}, \qquad \hat \beta_1 = \frac{\sum_{i=1}^m Y_i}{m} - \frac{\sum_{i=m+1}^n Y_i}{n-m}.
\end{equation*}
Convergence of sample quantities to population quantities follows from the law of large numbers; the continuous mapping theorem and standard probabilistic inequalities \citep{hansen2000} guarantee the ordinary least squares estimate is consistent: $(\hat \beta_0, \hat \beta_1) \arrowp (\beta_0, \beta_1)$.

We focus our attention on $\hat\beta_1$ as an estimator of the model parameter $\beta_1$. Consistency of $\hat\beta_1$ generally follows from the exogeneity conditions $\E[e_i] = 0$ and $\E[e_i X_i] = 0$. In this setting, $\hat\beta_1$ is not only consistent, but as a linear combination of two independent sample means, it has a number of desirable properties, including finite-sample unbiasedness, asymptotic normality, and root-$n$ consistency. Importantly, as a linear combination of two independent minimum variance unbiased estimates, $\hat \beta_1$ is the minimum variance unbiased estimate of $\beta_1$, such that, barring the use of auxiliary information, efficiency gains cannot be obtained without introducing the possibility of finite-sample bias. (The same results hold for $\hat \beta_0$ as an estimator of $\beta_0$.) This efficiency result holds for all ``saturated" linear models where the predictors are composed of noncollinear disjoint indicator variables, but does not generally hold for linear models with, e.g., continuous predictors. Consistency and asymptotic normality are preserved for any linear model such that the semiparametric identification conditions (generalized appropriately) hold.

\subsection{Standard Error Estimation}

An exact expression for the standard error of $\hat \beta_1$ can be obtained directly as a linear combination of independent random variables:
\begin{equation*} \label{se}
\sqrt{V(\hat\beta_1)} =\sqrt{\frac{\frac{1}{n-m} \sum_{i=m+1}^n \Var[e_i]} {n - m} + \frac{\frac{1}{m} \sum_{i=1}^m \Var[e_i]}{m}}.
\end{equation*}
This is the true standard error, and is derived purely in terms of our model and semiparametric identification conditions.

Let us consider robust standard error estimation. When we have mutual independence of the disturbances, robust standard error estimation generally proceeds by replacing each unobserved variance $\Var[e_i]$ with its empirical analogue, the squared residual $\hat e_i^2$. (The residual $\hat e_i = Y_i - \hat\beta_0 - \hat\beta_1 X_i$.) 
The robust standard error estimator of $\sqrt{V(\hat\beta_1)}$, $$
\sqrt{\hat V_{Het}(\hat\beta_1)} = \sqrt{\frac{\frac{1}{n-m} \sum_{i=m+1}^n \hat e_i^2} {n - m} + \frac{\frac{1}{m} \sum_{i=1}^m \hat e_i^2}{m}}.
$$
(See, e.g., \citet[Ch. 7]{imbensrubin} for a straightforward derivation.) Slutsky's Theorem ensures that each of the squared residuals is asymptotically unbiased for its observation's variance: $\E[\hat e_i^2] \rightarrow \E[e_i^2] = \Var[e_i], \forall i$. Thus, by the law of large numbers, both of the conditional average disturbance variances can be consistently estimated: $\frac{1}{n-m} \sum_{i=m+1}^n \hat e_i^2 - \frac{1}{n-m} \sum_{i=m+1}^n \Var[e_i] \arrowp 0$ and $\frac{1}{m} \sum_{i=1}^m \hat e_i^2 - \frac{1}{m} \sum_{i=1}^m \Var[e_i] \arrowp 0$. It follows that $\sqrt{\hat V_{Het}(\hat\beta_1)} $ is consistent, i.e.,
$
\sqrt{n \hat V_{Het}(\hat\beta_1)} - \sqrt{n V(\hat\beta_1)} \arrowp 0.
$

The classical standard error takes a different form. Classical standard error estimation is typically motivated by a homoskedasticity assumption: $\Var[e_i]=\Var[e_j], \forall i,j$. Note that if this were true, then $\Var[e_j] = \frac{1}{n} \sum_{i=1}^n \Var[e_i], \forall j$. Under homoskedasticity, instead of replacing each unobserved variance $\Var[e_i]$ with its empirical analogue $\hat e_i^2$, we could instead replace it with the {\it average} squared residual over all observations: $\frac{1}{n} \sum_{i=1}^n \hat e_i^2$. This motivates the classical standard error estimator (ignoring degree of freedom corrections):
$$
\sqrt{\hat V_{C}(\hat\beta_1)} = \sqrt{\frac{\frac{1}{n} \sum_{i=1}^n \hat e_i^2} {n - m} + \frac{\frac{1}{n} \sum_{i=1}^n \hat e_i^2}{m}}.
$$
The classical standard error attempts to estimate the conditional average disturbance variances by pooling across the dataset and computing the average squared residual, instead of estimating the conditional average disturbance variances separately. 

Although the average squared residual is consistent for the average disturbance variance ($\frac{1}{n} \sum_{i=1}^n \hat e_i^2 - \frac{1}{n} \sum_{i=1}^n \Var[e_i] \arrowp 0$), it is not generally the case that the average squared residual will be consistent for either conditional average disturbance variance (i.e., neither $\frac{1}{n} \sum_{i=1}^n \hat e_i^2 - \frac{1}{n-m} \sum_{i=m+1}^n \Var[e_i] \arrowp 0$ nor $\frac{1}{n} \sum_{i=1}^n \hat{e}_i^2 - \frac{1}{m} \sum_{i=1}^m \Var[e_i] \arrowp 0$). Thus, unlike the robust standard error, the classical standard error estimator is not generally consistent: 
$ 
\sqrt{n \hat V_{C}(\hat\beta_1)} - \sqrt{n V(\hat\beta_1)} \arrowp c \neq 0
$
 unless $m/n \rightarrow \mu = 0.5$ or $\frac{n}{n-m} \sum_{i=m+1}^n \Var[e_i] - \frac{n}{m} \sum_{i=1}^m \Var[e_i] \rightarrow 0$, as would be implied by homoskedasticity. Note that this result is a special (and simplified) case of \citet{white1980}'s Theorem 3.

The classical standard error in this setting corresponds to a consistent estimator under the assumption of a parametric submodel for which the ordinary least squares estimator is asymptotically efficient. Namely, it corresponds to the standard error when it is assumed that the data is generated by a normal-linear model with i.i.d. disturbances. The classical standard error estimator itself is robust to the selection of parametric models: it is consistent given any submodel of the semiparametric model such that the disturbances are mutually independent and homoskedastic. This in fact provides a semiparametric motivation for the classical standard error. In the specific case of our simple linear model with binary $X_i$, necessary and sufficient conditions for consistency of the classical standard error are given by $m/n \rightarrow \mu = 0.5$ or $\frac{n}{n-m} \sum_{i=m+1}^n \Var[e_i] - \frac{n}{m} \sum_{i=1}^m \Var[e_i] \rightarrow 0$. But without strong theoretical justification for this additional assumption, there is no general reason to believe that the classical standard error and the robust standard error will agree even asymptotically. When this assumption -- unrelated to the semiparametric modeling assumptions in Section 2.1 -- fails, the classical standard errors will not be consistent, and the robust standard errors will be consistent.

\subsection{Behavior under a Parametric Submodel}

Let us investigate the behavior of robust and classical standard error estimators when we have assumed that a particular parametric model holds that is compatible with the semiparametric assumptions. We now assume that the disturbance term is heteroskedastic and nonnormal. Specifically suppose that

$$e_i \sim \left\{
     \begin{array}{lr}
     Laplace(0,1) & : X_i = 0   \\
     Laplace(0,2) & : X_i = 1
     \end{array}\right..$$
Note that while there is heteroskedasticity ($\E[e_i^2 | X_i = 0] = 2 \neq 8 = \E[e_i^2 | X_i = 1]$), this does not contradict any of the model assumptions. All of the assumptions hold, and thus there is no model misspecification.

However, the classical standard error and the robust standard error may disagree even asymptotically. Substituting values into the formulas given in Section 2.3 and taking limits, we have 
\begin{align*}
\sqrt{n V(\hat\beta_1)} &\rightarrow \sqrt{ \frac{8}{\mu} + \frac{2}{1-\mu}} = \sqrt{ \frac{8-6\mu}{\mu(1-\mu)}},\\
\sqrt{n \hat V_{Het}(\hat\beta_1)} &\arrowp \sqrt{ \frac{8}{\mu} + \frac{2}{1-\mu}} = \sqrt{ \frac{8-6\mu}{\mu(1-\mu)}},\\
\sqrt{n \hat V_{C}(\hat\beta_1)} & \arrowp \sqrt{ \frac{8 \mu + 2(1-\mu)}{\mu} + \frac{8 \mu + 2(1-\mu)}{1-\mu}} = \sqrt{ \frac{2+6\mu}{\mu(1-\mu)}}.
\end{align*}
Note that this implies that $\sqrt{n \hat V_{C}(\hat\beta_1)} - \sqrt{n V(\hat\beta_1)} \arrowp 0$ if and only if $m/n \rightarrow \mu = 0.5$. Similarly, any consistent test against the null hypothesis of homoskedastic, normal errors will reject with probability one in large samples.\footnote{N.b., although some simulation evidence is presented, neither \citet{dhaenehoorelbeke2004} nor KR prove consistency of their proposed parametric bootstrap-based information test for any family of non-local alternatives.} 

Note that if the distribution of the disturbances were both nonnormal and known, the OLS estimate would not be asymptotically efficient. Rather, if the researcher knew that $e_i$ were Laplace distributed, the difference in medians would be the maximum likelihood estimate of $\beta_1$ and would have lower asymptotic MSE. However, the maximum likelihood estimate would not be robust to misspecification of the disturbance distribution: if incorrectly specified (i.e., for any true disturbance distribution that was asymmetric), the difference-in-medians estimator would not be consistent, much less efficient. In contrast, the semiparametric ordinary least squares estimator retains consistency without full specification, and the robust standard errors correctly reflect sampling variability.

\subsubsection{Monte Carlo Simulation}

To illustrate the behavior of robust and classical standard errors under the assumed parametric model in Section 2.4, we simulate data given $m=200$, $n=1000$. Figure 1 presents the sampling distributions of the robust standard errors and classical standard errors, computed using 25000 simulations.

\begin{figure}[ht]
\centering
\centerline{\includegraphics[width=4in]{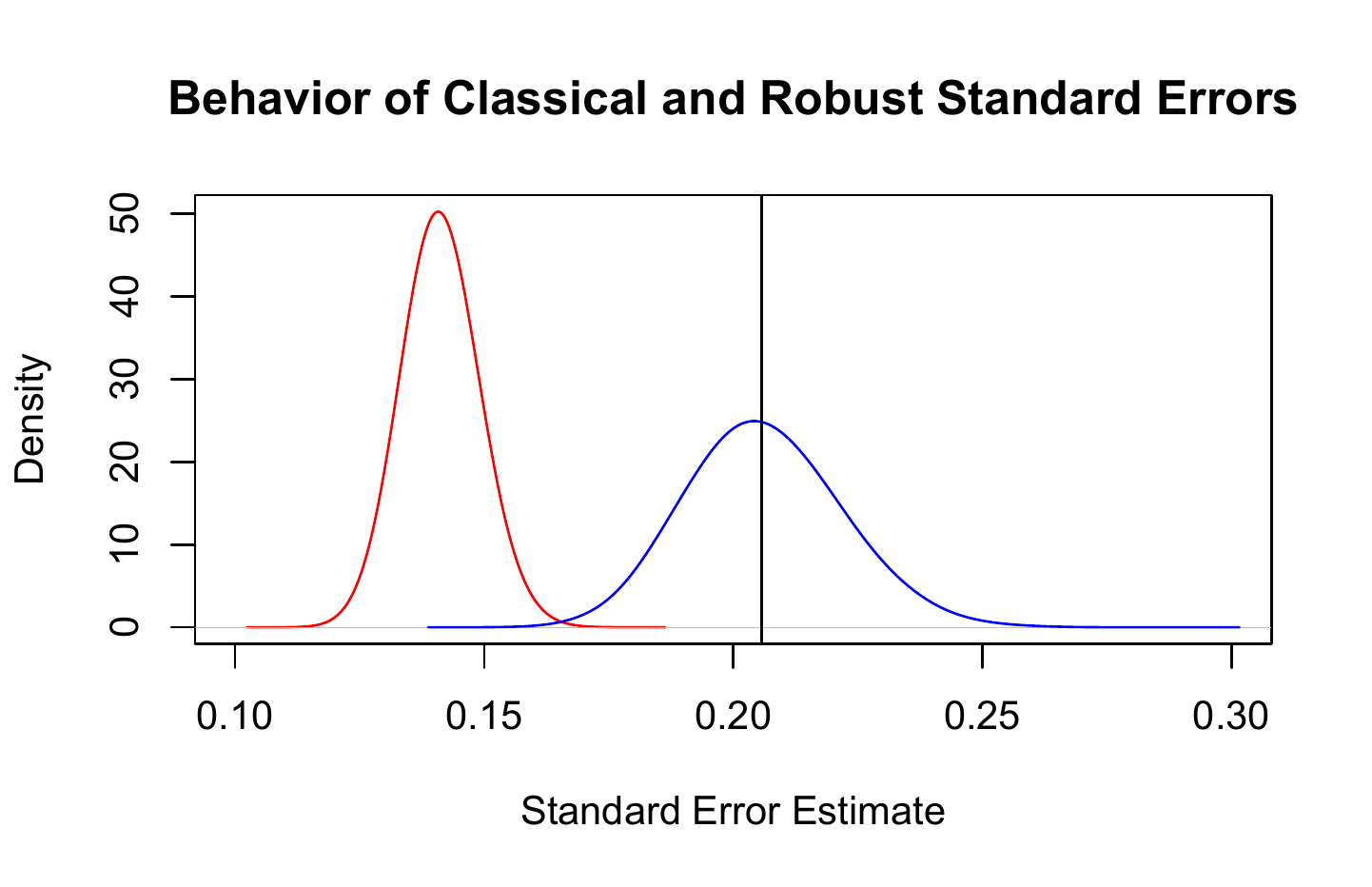}}
\caption{Illustration of the behavior of robust and classical standard errors, given the parametric model in Section 2.4, with $m=200$ and $n=1000$. The red curve represents a density estimate (from 25000 simulations) of the sampling distribution of the classical standard error for $\hat\beta_1$. The blue curve represents the density estimate (from 25000 simulations) of the sampling distribution of the robust standard error for $\hat\beta_1$. The black vertical line indicates the true standard error for $\hat\beta_1$.}
\end{figure}


The true standard error in this setting is $0.206$ and, as expected, the robust standard error estimator is approximately unbiased here ($\E[ \sqrt{\hat V_{Het}(\hat\beta_1)}] = 0.206$). The classical standard error estimator is severely biased in this setting ($\E[ \sqrt{\hat V_{C}(\hat\beta_1)}] = 0.141$), and this bias is not alleviated in large samples. Put simply: KR's proposed heuristic -- comparing the robust standard error to the classical standard error -- does worse than provide no information about misspecification in this setting: it provides misinformation. Why? The classical standard error relies on a modeling assumption that we {\it did not make} (namely, homoskedasticity) and is not true in this setting. The robust standard error accurately reflects the sampling variability of $\hat\beta_1$. 


\section{Discussion}

We have no aversion to parametric models when justified by the researcher. This is especially true in small samples, where additional structure can be used profitably for efficiency gains, even if the assumptions are incorrect. Similarly, the small sample behavior of robust standard errors \citep{angristpischke2009,imbenskolesar2015} may suggest measures of uncertainty that exploit more structure (e.g., classical standard errors) are preferable when $n$ is small. In such cases, a qualitative disagreement between the robust and classical standard errors may simply suggest high variability in the robust standard error, but would not speak to misspecification. The world of small samples is a difficult one -- filled with tradeoffs -- and we hesitate to make any general recommendations. 

But even given inquiry using parametric models, KR's recommendations are problematic. KR recommend the following procedure: ``If your robust and classical standard errors differ, follow venerable best practices by using well-known model diagnostics to evaluate and then to respecify your statistical model. If these procedures are successful, so that the model now fits the data and all available observable implications of the model specification are consistent with the facts, then classical and robust standard error estimates will be approximately the same" (p. 160). Compare this to the definition of data snooping from \citet{white2000}: ``Data snooping occurs when a given set of data is used more than once for purposes of inference or model selection. When such data reuse occurs, there is always the possibility that any satisfactory results obtained may simply be due to chance rather than to any merit inherent in the method yielding the results." To follow KR's recommendations without some accounting for the data-adaptive nature of the model selection may yield statements of confidence that are too strong -- both about the model parameters as well as the validity of the model itself. 

To summarize, KR's recommendations demand that we specify a full parametric model that reduces the phenomenological description of the observable data to a finite-dimensional function. KR then advocate that we readjust the model based on the data until we find a model that fits. That is, when the data does not match this infinitely restrictive model, we should change the assumptions until it does. And thus we conclude by evaluating KR's recommendations in light of \citet{manski2003}'s {\it Law of Decreasing Credibility}: ``The credibility of inference decreases with the strength of the assumptions maintained." Taking Manski's law at face value, then a semiparametric model is definitionally more credible than any assumed parametric submodel thereof. It is our view that the credibility of quantitative inquiry will be reduced if researchers are forced to adopt a full parametric model without strong ex ante theoretical justification. Thus we would suggest that researchers exercise a great deal of caution in adopting KR's recommendations.

\begingroup

\end{document}